# Experimental and Computational Investigation of Layer Dependent Thermal Conductivities and Interfacial Thermal Conductance of 1- to 3-Layer WSe2

*Elham Easy[1], Yuan Gao[2], Yingtao Wang[1], Dingkai Yan[3], Seyed M. Goushehgir[4], Eui-Hyeok Yang[1], Baoxing Xu[2]\*, Xian Zhang[1]\**

[1]Department of Mechanical Engineering, [3]Department of Chemical Engineering and Materials Science, Stevens Institute of Technology, Hoboken, New Jersey 07030, United States

[2]Department of Mechanical and Aerospace Engineering, University of Virginia, Charlottesville, Virginia 22904, United States

[4]Department of Mechanical Engineering, Urmia University of Technology, Urmia, West Azerbaijan, Iran

ABSTRACT. The two dimensional materials such as graphene and transition metal dichalcogenides (TMDC) have received extensive research interests and investigations in the past decade. In this research, we used a refined opto-thermal Raman technique to explore the thermal transport properties of one popular TMDC material $WSe_2$, in the single layer (1L), bi-layer (2L), and tri-layer (3L) forms. This measurement technique is direct without additional processing to the material, and $WSe_2$'s absorption coefficient is discovered during the measurement process to further increase this technique's precision. By comparing the sample's Raman spectroscopy spectra through two different laser spot sizes, we are able to obtain the two parameters - lateral thermal conductivities of 1L-3L $WSe_2$, the interfacial thermal conductance between 1L-3L $WSe_2$ and the substrate. We also implemented full-atom nonequilibrium molecular dynamics simulations (NEMD) to computationally investigate the thermal conductivities of 1L-3L $WSe_2$ to provide comprehensive evidence and confirm the experimental results. The trend of layer dependent lateral thermal conductivity and interfacial thermal conductance of 1L-3L $WSe_2$ is discovered. The room temperature thermal conductivities for 1L-3L $WSe_2$ are 37 ± 12, 24 ± 12, and 20 ± 6 W/(m·K), respectively. The suspended 1L $WSe_2$ possesses a thermal conductivity of 49 ± 14 W/(m·K). Crucially, the interfacial thermal conductance between 1L-3L $WSe_2$ and the substrate is found to be 2.95 ± 0.46, 3.45 ± 0.50, and 3.46 ± 0.45 MW/(m$^2$·K), with a flattened trend starting the 2L, a finding that provides the key information for thermal management and thermoelectric designs.

TEXT.

INTRODUCTION

Since graphene's first isolation by mechanical exfoliation in 2004 [1-3], two dimensional (2D) materials have received extensive attention due to their unique atomically thin structure and novel physical properties [4-23]. In particular, transition metal dichalcogenides (TMDC) materials have shown intriguing thermal, thermoelectric, electrical and optical properties, such as enhanced figure of merit, prominent band structure, and semiconducting behavior, which distinguishes them from graphene [8-10, 24-33]. This makes TMDC materials an ideal candidate for the next-generation thermal, thermoelectric, electrical, and optical applications. Although the common TMDC materials - 2D $MoS_2$ and $MoSe_2$'s thermal properties have been studied extensively both experimentally and theoretically [30, 32, 34-43], the investigation of the emerging thermoelectric material $WSe_2$ has received much less attention. On the other hand, 2D $WSe_2$ has emerged as a promising thermoelectric candidate due to the high Seebeck coefficient of 680 μV/K [69], compared to 280 μV/K for $Bi_2Te_3$ - a commercial thermoelectric material. There is an increasing layer dependent trend of Seebeck coefficient and decreasing trend of thermal conductivity of $WSe_2$, leading to a high figure of merit with the layer number. Thus discovering the thermal conductivities of 2D $WSe_2$ and the layer dependent trend is of fundamental importance and also lays immediate guides to the development of flexible thin film thermoelectric devices, thereby revolutionizing both fundamental thermal transport of materials and application technologies in thermoelectric devices.

While the electrical and optical properties of 2D $WSe_2$ have been extensively studied [44-51], there is no experimental research reported on thermal transport properties in 2D $WSe_2$. The computational studies of $WSe_2$'s thermal transport properties have the results in the large range of 0.2 - 22 W/(m·K) [52-57] which motivates experimental investigation and verification. Meanwhile, a low thermal conductivity in the range of 1.2-1.6 W/(m·K) was found for disordered layered $WSe_2$ in the bulk form by both electrical heating method and time-domain thermoreflectance (TDTR) method [58, 59]. In a recent thermal measurement based on TDTR, a thermal conductivity value of 42 W/(m·K) was discovered for the ordered crystalline layered $WSe_2$ in the bulk form [60]. Therefore, it is needed to conduct a comprehensive experimental thermal investigation of $WSe_2$ in the 2D atomic-layered forms (single-layer, bi-layer, and tri-layer) to discover the intrinsic in-plane thermal conductivities and the interfacial thermal conductance of 2D $WSe_2$, and investigate their layer dependence.

In this letter, we demonstrate the measurement of thermal conductivities of single-layer (1L), bi-layer (2L), and tri-layer (3L) $WSe_2$ and their interfacial thermal conductance to the substrate, via a well-developed opto-thermal Raman technique for discovering thermal properties of TMDCs in the 2D form. We also carefully measured 1L-3L $WSe_2$'s absorption coefficients during the measurement process. The systematic experimental results provide a layer number dependent analysis. The opto-thermal

Raman technique has been the most suitable method for measuring thermal conductivity of 2D materials down to sub-nanometer thickness [20, 30, 37]. Compared to TDTR method which is famous for its capability of measuring the thermal conductivities of 2D materials in the bulk form [61] and electrical heating method [59] which is renowned for its reliability in high-resolution temperature calibration for 2D bulk materials, opto-thermal Raman technique is fit for measuring intrinsic thermal conductivities of 2D materials with thickness down to sub-nanometer. In opto-thermal Raman technique, laser is focused on the sample and the position of the Raman peaks are utilized as the benchmark. As the laser power is increased, sample is heated, which enables red-shift Raman mode due to thermal softening. Thermal conductivity is then calculated from the thermal modeling based on the measured Raman position shift rate. In addition, the thermal modeling requires the input of several key parameters: Raman peak position shift with temperature, optical absorbance, and the interfacial thermal conductance between $WSe_2$ and the substrate. The retrieved room temperature thermal conductivities and interfacial thermal conductance of the 1L-3L $WSe_2$ are the first reported experimental results. In summary, opto-thermal Raman technique based measurement is conducted at steady state and independent of time, during which a Raman laser is focused on the sample surface and used as both the heat source and a temperature prober. As the heat source, laser power is absorbed uniformly throughout the thickness, making it an ideal method in measuring thermal conductivities of samples with the smallest existing thickness existing (sub-nanometer). In addition, the non-contact measurement method has the minimum fabrication requirement and thus maintains the sample's pristine physical properties.

RESULTS AND DISCUSSION

Figure 1a presents the schematics of WSe$_2$ on SiO$_2$/Si substrate with the opto-thermal Raman measurement technique by a 532 nm wavelength laser. Figure 1b shows the laser profile and thermal transport schematics in WSe$_2$. Figure 1c shows the atomic force microscopy (AFM) topography of 1L, 2L and 3L WSe$_2$ flakes obtained by mechanical exfoliation, with the optical microscope image in Figure 1d as a reference. Mechanical exfoliation method has been a key to provide high-quality 2D materials that ensures pristine physical properties of the sample. The thickness of 1L-3L WSe$_2$ samples are confirmed by AFM. The detailed AFM information is recorded in Supporting Information. Figure 1e presents a scanning electron microscopy (SEM) image of the suspended 1L WSe$_2$ sample on 1 μm and 1.5 μm diameter holes. Figure 1f demonstrates the Raman spectrum of WSe$_2$ by 532 nm wavelength laser, with $E^1_{2g}$ and $A_{1g}$ peaks. $E^1_{2g}$ peak is used as the characterization peak for thermal transport measurement due to its clear and accurate signal. Figure 1g shows photoluminescence spectra of 1L-3L by 532 nm wavelength laser, which also reconfirms the 1L WSe$_2$ because it has a direct bandgap which is presented as a single peak photoluminescence (PL) curve. 1L WSe$_2$ has only one peak at about 1.65 eV, while 2L and 3L WSe$_2$ has the PL peak at around 1.63 eV. The peak intensity of 1L WSe$_2$ at 1.65 eV is at least 2.5 times of that of 2L WSe$_2$ at 1.63 eV, in accordance with previous report [33]. PL curves also show the suspended WSe2 has a PL peak intensity that is more than 500 times than the supported WSe2 which reconfirms the suspended condition.

During the measurement process, a Raman laser with a wavelength of 532 nm is focused on WSe2 through the microscope objective, and the thermal flow provided by the laser is propagating isotropically in all radial directions from the laser position. Here we are considering measured temperature, laser spot size, absorption coefficient of the sample, and laser power. The 40 × and 100 × microscope objectives were used to generate two laser spot sizes, providing enough experimental data to obtain the two key parameters – in-plane thermal conductivity, and interfacial thermal conductance. To determine the heat source which is the laser power absorbed for the thermal transport analysis, another group of 1L-3L WSe2 from the same crystal source were prepared on the transparent quartz substrates, and their optical absorption spectra were measured to determine the frequency dependent complex dielectric function. The dielectric functions of the target sample material and that of the substrate material, were used to calculate the absorption coefficient at the wavelength of 532 nm based on the standard transfer matrix method. The absorption coefficient $\alpha$ of WSe2 can be determined by the optical transmittance $T_e$ taken from the measurements and calculated based on the following equation [62]:

$$T_e = \frac{(1-R)^2 exp(-\alpha d)}{1-R^2 exp(-2\alpha d)} \tag{1}$$

, where $d$ is the target sample thickness and $R$ is a function of the refractive index $n$: $R = (n-1/n+1)^2$, with $n$ reconfirmed from [29]. The obtained $\alpha$ is 894734 cm$^{-1}$ at the wavelength of 532 nm. The obtained optical absorbance of 1L-3L WSe$_2$ are 5.7 ± 1.1 %, 10.9 ± 2.1 %, and 16.0 ± 2.8% at 532 nm wavelength. These values can be used for the suspended WSe$_2$ sample. For the supported WSe$_2$, the optical interference effect from the substrate (285 nm SiO$_2$ on Si) was taken into account to calculate the absorbance, with results summarized in Table 1. Because of the low quantum yield of as-exfoliated 1L WSe$_2$ [70-73] which is in the range of 0.03% - 3%, the absorbed energy emitting as photons is ignored.

The laser beam spot ($r_0$) is characterized by moving the laser across a sharp edge based on that the Raman laser profile is following a Gaussian distribution. Figure 2a shows the Raman spectroscopy mapping across a sharp edge to monitor the Si peak. The mapping frequency range is 500 cm$^{-1}$ - 550 cm$^{-1}$ to monitor the Si peak. Figure 2b, c shows optical microscope image and 10 μm × 10 μm micro-Raman mapping. The Si peak intensity was studied as a function of the moving position ($dI/dx$) with Gaussian distribution ($K \cdot exp(-(x-a)^2/r_0^2)$) fitting (Figure 2d, e). Through characterization and fitting, the laser spot size passing through the 100 × objective is determined as 0.18 ± 0.02 μm and that passing through the 40 × objective is determined as 0.26 ± 0.02 μm. $r_0$ can also be estimated from the numerical aperture estimation: $r_0 = \lambda/(\pi \cdot NA)$. $NA$ is the numerical aperture value of the objectives, and is 0.9 and 0.75 for 100 × and 40 × objectives. $r_0$ was calculated as 0.19 μm and 0.23 μm. This estimated value is close to the experiment results.

Raman laser was used as both the heat source and the temperature characterization. We first calibrated the Raman $E^1_{2g}$ peak position's shift rate with the temperature ranging from 298 K to 473 K. A laser with wavelength $\lambda$ of 532 nm was used. Figure 3a presents the Raman spectra of 1L WSe$_2$ with temperature dependence. Figure 3b shows the temperature dependent Raman $E^1_{2g}$ peak position shift of 1L WSe$_2$ in both supported and suspended forms. The peak position's linear red shift is caused by the bond softening that's thermally driven, which is consistent with the previous studies on graphene, MoS$_2$ and MoSe$_2$ [30]. The suspended sample has a larger thermally driven bond softening due to the larger in-plane lattice expansion in the suspended condition. Figure 3c, d presents the temperature dependent of the $E^1_{2g}$ peak shift for 2L, 3L WSe$_2$, showing the same red shift. The linear temperature dependent shift rate of the $E^1_{2g}$ peak is defined as the first-order temperature coefficient (Table 1). It is found that the first-order temperature coefficient decreases with layer number, and this decreasing trend is consistent with the previous work on MoS$_2$ and MoSe$_2$ [30], due to the fact that with the increasing layer number of 2D WSe$_2$, the temperature dependent in-plane lattice expansion coefficient decreases.

In the second part of the thermal measurements, laser power dependent Raman spectra were taken to extract the linear relation between sample temperature and absorbed laser power. The $E^1_{2g}$ Raman peak shift was fit as a function of the absorbed laser power for 1L-3L WSe$_2$ samples, and two laser spots with size of 0.18 and 0.26 μm were used. Figure 3e-g present the laser power dependent Raman peak shift rates of the supported 1L-3L WSe$_2$ and suspended 1L WSe$_2$. The obtained absorbance, temperature coefficients, and absorbed laser power shift rates of 1L-3L WSe₂ are summarized in Table 1.

The schematics of laser heating process at steady state when WSe₂ is shown in Figure 1a, b. The temperature distribution in cylindrical coordinate $T(r)$ is related to the absorbed laser power, in-plane thermal conductivity and interfacial thermal conductance between sample and substrate [30]. The heat convection from the WSe₂ to the air is less than 0.01% of the total heat in the thermal transport process and is thus ignored (see Supporting Information). The absorbed laser power $P$ is determined by the multiplication of the actual laser power and the sample's optical absorbance (Table 1). Based on the Gaussian profile of laser and its spot size r₀, the volumetric laser heating power density q‴(r) is presented as:

$$q'''(r) = P \cdot \frac{1}{t} \cdot \frac{1}{\pi r_0^2} \exp\left(-\frac{r^2}{r_0^2}\right) \tag{2}$$

, where $t$ is the thickness of WSe₂. Considering the interfacial thermal conductance between WSe₂ and substrate, $T(r)$ is expressed as:

$$\frac{1}{r}\frac{d}{dr}\left(r\frac{dT(r)}{dr}\right) - \frac{g}{k_s t}(T(r) - T_a) + \frac{q'''(r)}{k_s} = 0 \tag{3}$$

Here $T_a$ is the substrate temperature, $\kappa_s$ is the in-plane thermal conductivity of WSe₂ in supported condition, and $g$ is the interfacial thermal conductance between WSe₂ and the substrate. COMSOL Multiphysics simulations have shown the substrate temperature $T_a$ also increases with laser heating, and the ratio between the temperature increase of WSe₂ and SiO₂ is used for obtaining the interfacial thermal conductance $g$ (see Supporting Information). The boundary conditions $(dT)/(dr)|_{r=0} = 0$ for laser's Gaussian profile and $T(r \to \infty) = 0$ for room temperature at the edge of the sample are applied. For the suspended sample, $g = 0$.

The average temperature of WSe$_2$ within the laser spot is characterized by the E$^1_{2g}$ peak position using the Raman peak shift rate in Table 1. The average temperature is a weighted result by considering the Gaussian profile of the laser spot:

$$T_m = \frac{\int_0^\infty T(r) \exp\left(-\frac{r^2}{r_0^2}\right) r \, dr}{\int_0^\infty \exp\left(-\frac{r^2}{r_0^2}\right) r \, dr} \tag{4}$$

The fitted Raman peak shift rates from the experimental results in Figure 3 are summarized in Table 1. The measured thermal resistance of WSe$_2$ can be calculated as $R_m = T_m/P$. To calculate the WSe$_2$ thermal conductivity $\kappa_s$ and interfacial thermal conductance to the substrate g, we followed the method of Cai et al. [37] Equations 2-4 provide an expression of R$_m$ as a function of $\kappa_s$ and g. Specifically, the ratio of two measured R$_m$ based on the two laser spot sizes is a function of g/$\kappa_s$. So the ratio of two R$_m$ is used to obtain g/$\kappa_s$. And the R$_m$ value for either of the two spot sizes can be used to extract the values of $\kappa_s$ and g. This calculation process provides $\kappa_s = 37 \pm 12$ W/(m·K) and g = $2.95 \pm 0.45$ MW/(m$^2$·K) for the 1L WSe$_2$, based on the measurements using 100× and 40× lenses with spot sizes of 0.18 μm and 0.26 μm. Using this calculation method, the thermal conductivities and interfacial thermal conductance to the substrate for the 2L - 3L WSe$_2$ and the suspended 1L WSe$_2$ are obtained and summarized in Table 2 and Figure 4a, b.

The $\kappa_s$ values of several layers of $WSe_2$ in our experimental studies are within the distribution range being approximately 20-37 W/(m·K), which is within the range of published simulation results acquired by combining the first-principles calculations with Boltzmann transport equation [63]. The thermal conductivity of $WSe_2$ is smaller than the other two common TMDC materials $MoS_2$, $MoSe_2$ [30]. This is due to the increased atomic mass in $WSe_2$ [63]. The thermal conductivity of $WSe_2$ also decreases with the layer number (Figure 4a) and the supported $WSe_2$ has a lower thermal conductivity than the suspended $WSe_2$. This decreasing trend of thermal conductivity was also discovered in the exfoliated hBN, $MoS_2$, and $MoSe_2$ [23, 30, 34] (Figure 4c) and is attributed to the intrinsic scattering mechanism: Umklapp phonon scattering due to the crystal anharmonicity of phonons in thicker samples [64]. In addition, the phonon mean free path $l$ and the relative contribution from ballistic resistance decreases with increasing layer.

Computational investigations were conducted in parallel to confirm the experimental investigations of $WSe_2$'s thermal conductivity and its layer dependent trend. Full-atom nonequilibrium molecular dynamics simulations (NEMD) were implemented on the atomistic model depicted in Figure 5a, where 1L and 2 L $WSe_2$ with a length (in x-direction) of 14.2 nm and a width (in y-direction) of 8.2 nm in AA stacking order are supported by an amorphous $SiO_2$ substrate. The heat flow was introduced in the x-direction to measure the thermal conductivity. Figure 5b demonstrates the 1L and 2L $WSe_2$'s thermal conductivities, which is obtained with the heat flux, the overall temperature gradient along the heat flow direction (inset), and the number of layers. For WSe$_2$, a decrease in the thermal conductivity is observed as the layer number increases from 1 to 2, which is consistent with the measurements. To investigate the phonon mechanisms, the vibrational density of states (VDOS) of the supported WSe$_2$ were calculated and presented in Figure 5c. Compared to the VDOS of the 1L WSe$_2$, the VDOS of 2L WSe$_2$ possesses two suppressed peaks at 5.2 and 8.2 THz, which indicates more severe phonon scattering, weaker phonon resonance, and thus degraded thermal conductivity. Figure 5d confirms the robustness of the decrease in thermal conductivity with 4 different stacking orders, namely AA, AB', AA' and AB. The experimental measurement which is based on naturally exfoliated 2D WSe$_2$ is AA' stacking. The computational process is detailed in Supporting Information.

The computational investigation has confirmed the thermal conductivities of 2D WSe$_2$ and the decreasing trend with layer number, with proves more severe phonon scattering from 1L to 2L WSe$_2$ from the view of vibrational density of states. To further layer dependent trend of thermal conductivities of the broad family of 2D materials, we provided the existing published studies of 2D materials' layer dependent thermal conductivities in Table 3. It has been discovered that in the 2D regime, the exfoliated 2D materials' thermal conductivities have a decreasing trend with layer number [23, 30, 34], whereas the chemical vapor deposition (CVD) synthesized 2D materials' thermal conductivities have an increasing trend with layer number [66, 67]. Based on our computational investigation, this interesting opposite trend between exfoliated 2D materials and CVD 2D materials are due to the fact that exfoliated 2D materials are pristine. There are no grain boundaries in the exfoliated 2D materials, thus the Umklapp phonon scattering is dominant in the layer dependent trend and the total thermal conductivity is limited by the phonon Umklapp scattering and edge boundary scattering. As a result, the 2D materials with a few layers possess thermal conductivities that are significantly higher than that of the bulk materials which is in the range of 1 W/(m·K) [55]. Whereas in the CVD 2D materials, there are grain boundaries that prevent the efficient thermal transport, thus increasing layer number motivates more effective thermal transport paths.

EXPERIMENTAL SECTION

The mechanical exfoliation method was used to produce 1L, 2L, and 3L $WSe_2$ from the bulk crystal (SPI supplies). The substrate is $SiO_2$ (285 nm)/Si. The suspended substrate is also the same substrate with hole depth of 1 μm. The layer number of the $WSe_2$ sample was characterized by the tapping mode of atomic force microscopy (AFM) (Bruker Dimension), Raman spectroscopy and photoluminescence [33]. A micro Raman spectrometer (RENISHAW InVia Raman Microscope system) is used for the thermal transport measurements. $WSe_2$ on $SiO_2$/Si substrate was heated uniformly from 300 K to 500 K by a temperature controlled heating platform (Linkam Stage THMS600).

CONCLUSION

This work is the first experimental work about the in-plane thermal conductivities and interfacial thermal conductance of 1L - 3L $WSe_2$, whereas all the previous thermal studies of 2D $WSe_2$ are computational [52-57]. Our experimental method implemented an independent characterization of material absorbance, spot size, and studied interfacial thermal conductance, a critical parameter that determines the energy dissipation in electronic devices which ensures the precision of the measured value. These experimental results, along with the layer dependent trend of thermal conductivities are confirmed by our computational investigation using full-atom nonequilibrium molecular dynamics simulations. Our investigation of 1L-3L $WSe_2$'s thermal properties has also addressed several important problems in the general thermal study of 2D materials with atomic thickness. First, the decreasing trend of layer dependent in-plane thermal conductivity and interfacial thermal conductance of exfoliated $WSe_2$ are investigated both experimentally and computationally and a trend approaching a constant is observed on the sample of 3 layers. This trend has been compared to CVD materials with analysis. Second, it confirms the theoretical prediction that $WSe_2$ has a relatively low in-plane thermal conductivity among the family of 2D materials which proves it is a potential candidate for thermoelectrics. In general, we used the refined opto-thermal Raman technique to perform the first experimental study of thermal transport properties of 1L-3L $WSe_2$ with computational confirmation. Specifically careful characterizations of optical absorption coefficient, laser spot size, and thermal coupling to the substrate are conducted, which ensures a more robust thermal transport measurement in 2D materials. This work shed light on device modeling, thermal management, and quantum devices with enhanced performance.

FIGURES.

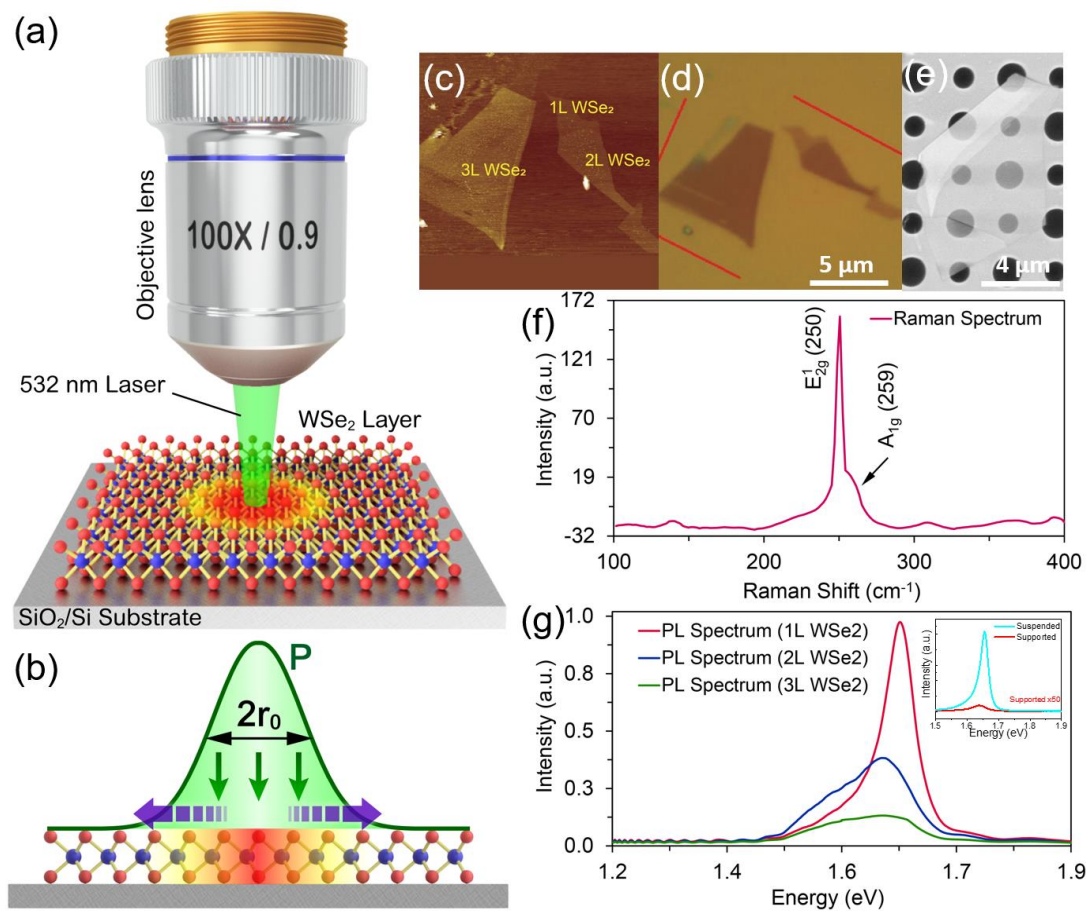

**Figure 1.** (a) Schematic of the experimental setup for the WSe₂ sample by opto-thermal Raman technique. (b) Schematic of Raman laser and thermal transport profiles. (c) Atomic force microscopy image of 1L-3L WSe₂. (d) Optical microscope image of 1L-3L WSe₂. (e) Scanning electron microscopy image of the suspended 1L WSe₂ on the holes. (f) Raman spectrum of 2D WSe₂ with two characteristic peaks ($E^1_{2g}$ and $A_{1g}$). (g) Photoluminescence spectra of 1L-3L WSe₂. Inset: Photoluminescence spectra of the supported and suspended areas specifically.

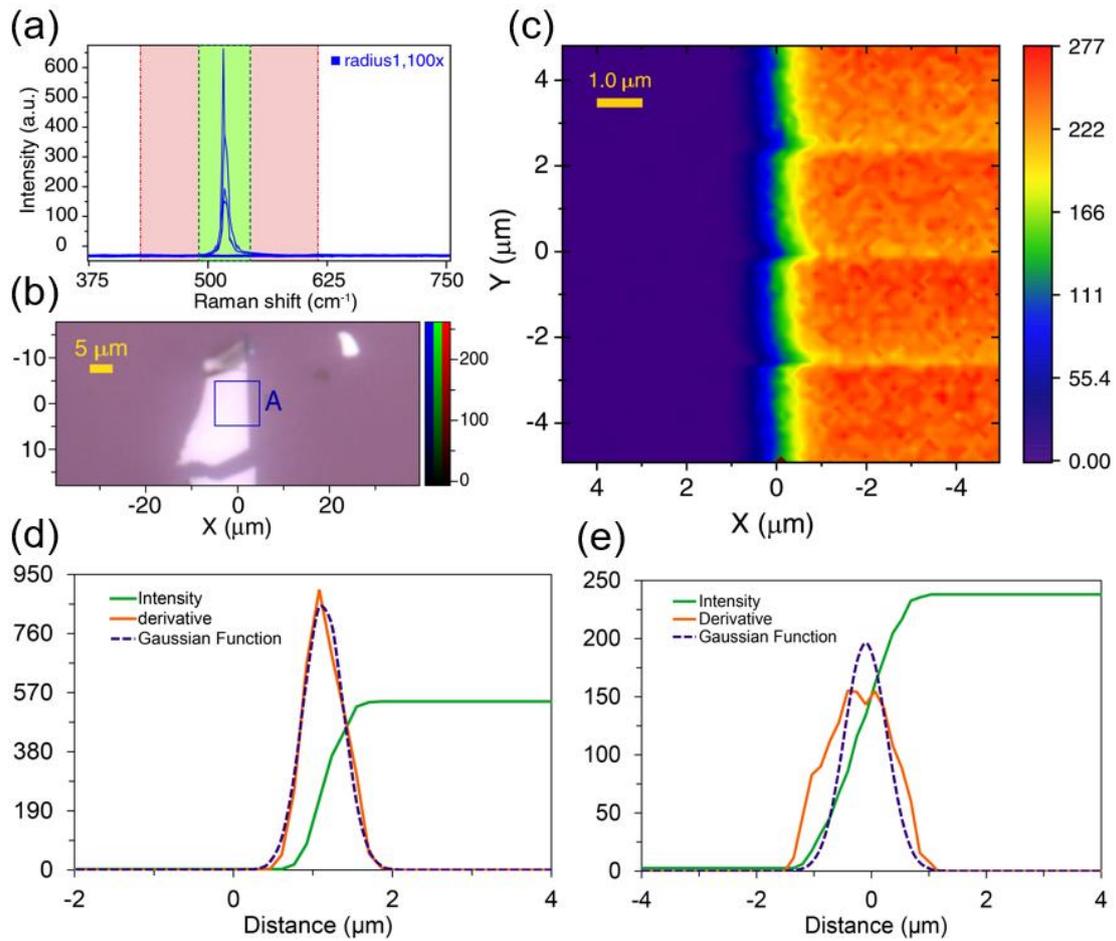

**Figure 2.** (a) The Raman spectra mapping monitoring Si peak. (b) Optical microscope image and (c) 10 μm × 10 μm micro-Raman mapping across a sharp flake edge. The Raman intensity (green) and extracted profile of the laser beam (orange) as a function of the beam position for laser spot from 100 × objective (d) and that from 40 × objective laser spot (e).

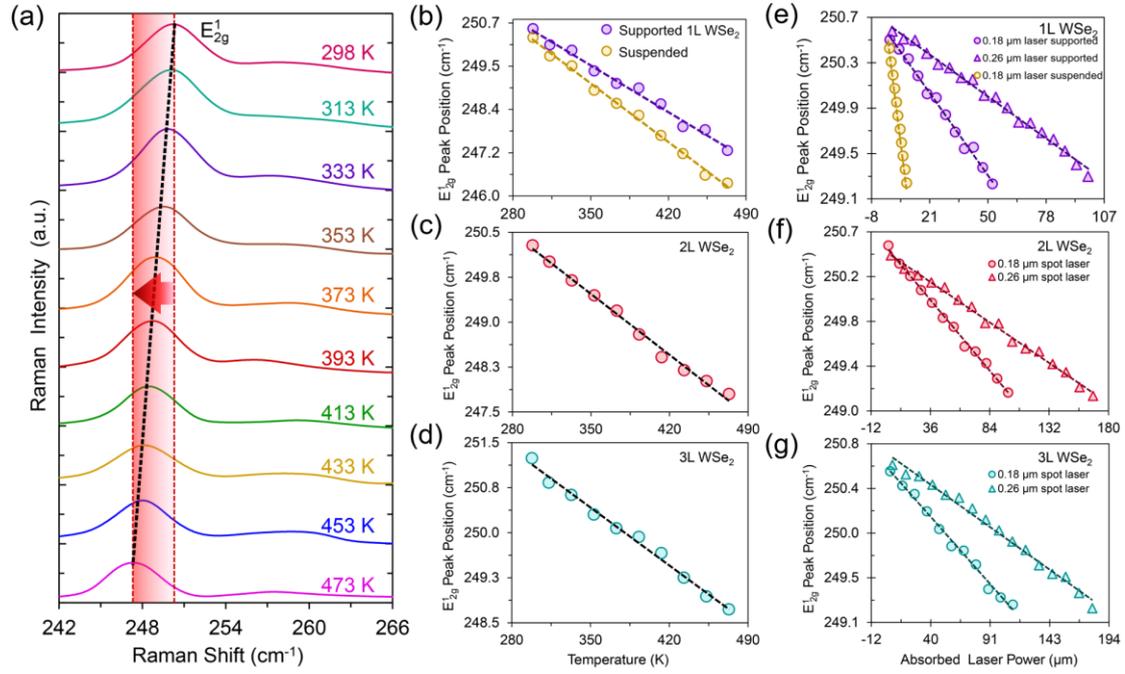

**Figure 3.** (a) Raman spectra of 1L WSe₂ recorded at temperatures from 298 K to 473 K. The temperature dependent $E^1_{2g}$ Raman peak shift measured on the 1L WSe₂ both supported and suspended (b), 2L WSe₂ (c), and 3L WSe₂ (d). Power dependent $E^1_{2g}$ Raman peak shift measured using different laser spot sizes, on the 1L WSe₂ both supported and suspended (e), 2L WSe₂ (f), and 3L WSe₂ (g).

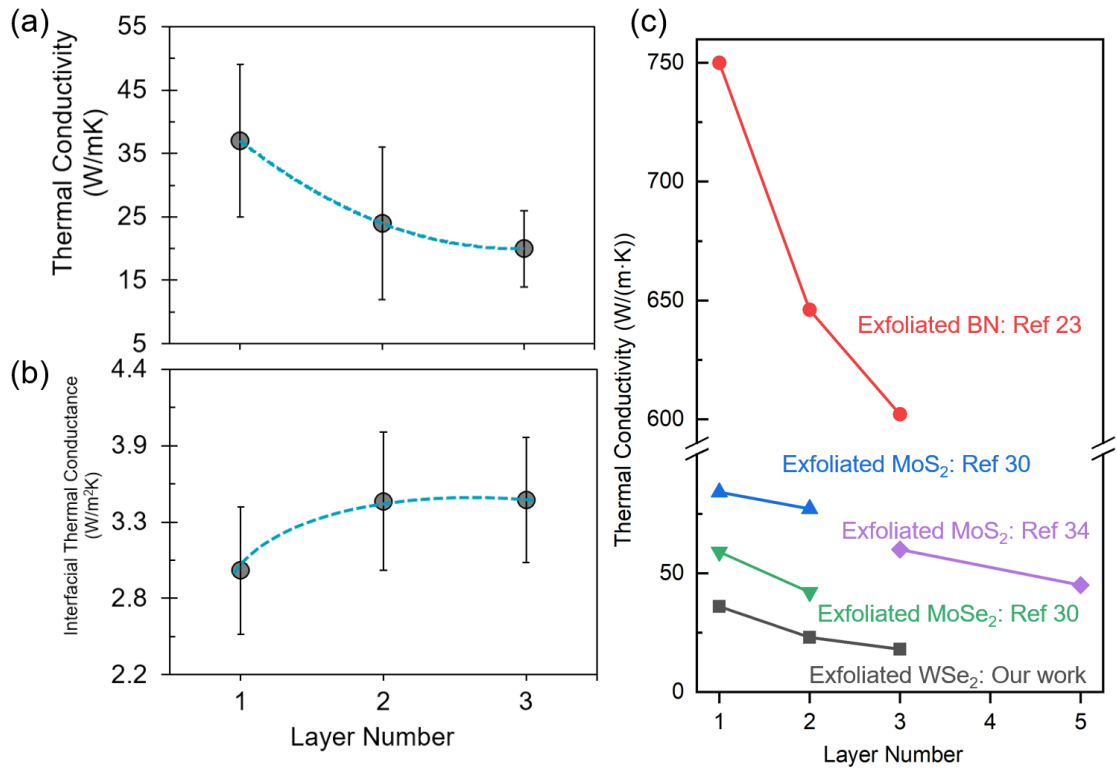

**Figure 4.** (a) Layer dependent in-plane thermal conductivity and (b) interfacial thermal conductance of WSe$_2$ from 1L to 3L WSe$_2$. (c) Comparison of layer dependent thermal conductivity of common exfoliated 2D materials [23, 30, 34].

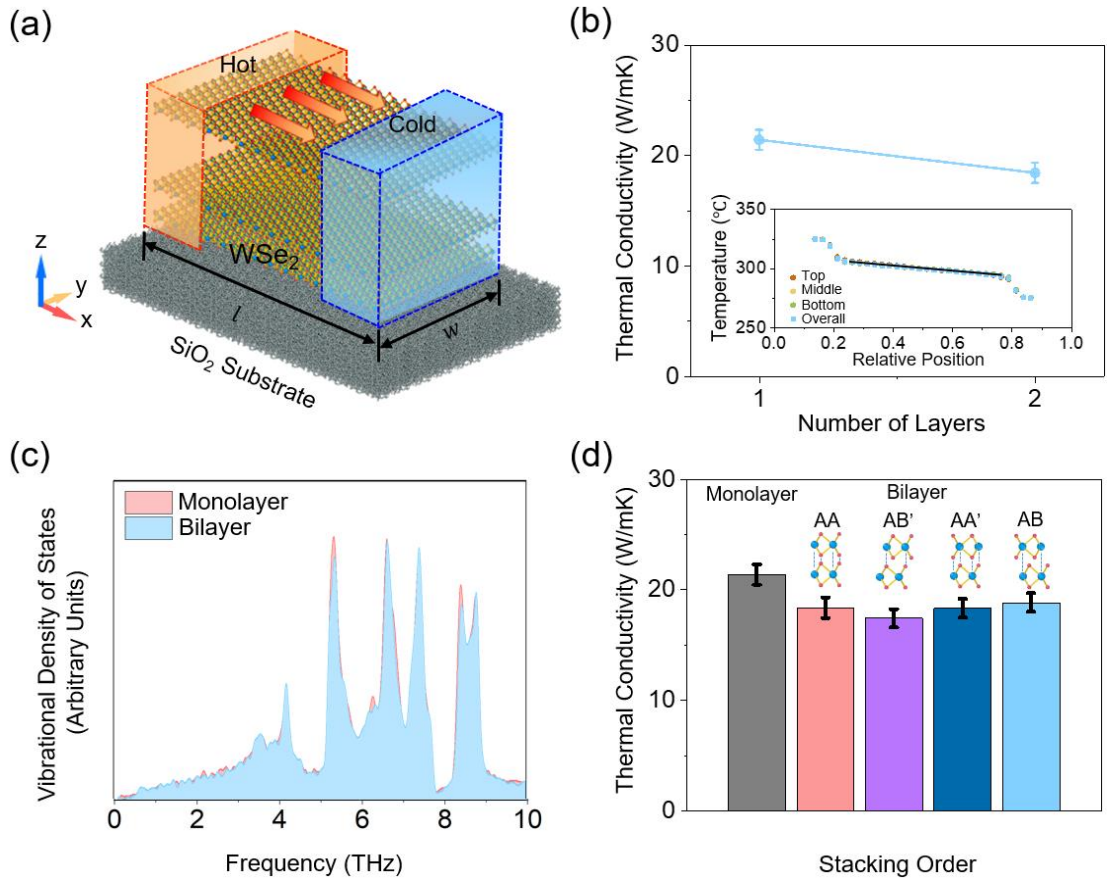

**Figure 5.** (a) Schematic of the atomistic modeling. (b) Thermal conductivity of WSe$_2$ as functions of the number of layers. (c) Vibrational density of states of 1L and 2L WSe$_2$. (d) Thermal conductivity of 2L WSe$_2$ with different stacking orders.

TABLES.

**Table 1.** First-order Temperature Coefficients, Absorbance, and Power Shift Rates of

1L-3L WSe$_2$

| | | Temperature coefficient (cm$^{-1}$/K) | Absorbance (%) | Absorbed power shift rate (cm$^{-1}$/μW) | |
| --- | --- | --- | --- | --- | --- |
| | | | | 0.18 μm spot | 0.26 μm spot |
| 1L WSe$_2$ | Supported | -0.0187 ± 0.0022 | 5.8 ± 1.1 | -0.0225 ± 0.0024 | -0.0124 ± 0.0150 |
| 1L WSe$_2$ | Suspended | -0.0226 ± 0.0025 | 5.7 ± 1.1 | -0.1385 ± 0.0129 | |
| 2L WSe$_2$ | Supported | -0.0147 ± 0.0019 | 11.1 ± 2.1 | -0.0142 ± 0.0028 | -0.0080 ± 0.0123 |
| 3L WSe$_2$ | Supported | -0.0132 ± 0.0004 | 16.3 ± 2.8 | -0.0125 ± 0.0004 | -0.0071 ± 0.0002 |

**Table 2.** Thermal Conductivities and Interfacial Thermal Conductance of 1L-3L WSe$_2$

| | | Thermal conductivity (W/(m·K)) | Interfacial thermal conductance (MW/(m$^2$·K)) |
| --- | --- | --- | --- |
| 1L WSe$_2$ | Supported | 37 ± 12 | 2.95 ± 0.46 |
| 1L WSe$_2$ | Suspended | 49 ± 14 | |
| 2L WSe$_2$ | Supported | 24 ± 12 | 3.45 ± 0.50 |
| 3L WSe$_2$ | Supported | 20 ± 6 | 3.46 ± 0.45 |

**Table 3.** Comparison of Layer-dependent Thermal Conductivity of Common 2D Materials (exfoliated 2D materials and CVD 2D materials with opposite trend)

| Material | Trend | Thermal conductivity at RT (W/(m·K)) | Reference |
|---|---|---|---|
| Exfoliated BN (1L-3L) | decreasing | 1L: 750; 2L: 646; 3L: 602 | 23 |
| Exfoliated $MoS_2$ (1L-2L) | decreasing | 1L: 84; 2L: 77 | 30 |
| Exfoliated $MoS_2$ (3L-5L) | decreasing | 3L: 60; 5L: 45 | 34 |
| Exfoliated $MoSe_2$ (1L-2L) | decreasing | 1L: 59; 2L: 42 | 30 |
| Computational studies of $MoS_2$ (1L-3L) | decreasing | 1L: 155; 2L: 125; 3L: 115 | 65 |
| CVD $MoS_2$ (1L-2L) | increasing | 1L: 13.3; 2L: 43.4 | 66 |
| CVD $WS_2$ (1L-2L) | increasing | 1L: 32; 2L: 53 | 67 |

## ASSOCIATED CONTENT

**Supporting Information**. The Supporting Information is available free of charge on the ACS Publications website.

Atomic force microscopy measurement of 1L-3L $WSe_2$, convection from the suspended material to the air, computational modeling and methodology (PDF).

## AUTHOR INFORMATION

**Corresponding Authors**

*Baoxing Xu, Email: bx4c@virginia.edu

*Xian Zhang. Email: xzhang4@stevens.edu

**Author Contributions**

The manuscript was written through contributions of all authors. All authors have given approval to the final version of the manuscript.

**Funding Sources**

Stevens Institute of Technology's startup funding, Office of Naval Research Young Investigator Program (grant number N00014-20-1-2611), Stevens Institute of Technology's Bridging Award, and Brookhaven National Laboratory Center for Functional Nanomaterials.

ACKNOWLEDGMENT

This work was supported by Stevens Institute of Technology's startup funding, Stevens Institute of Technology's Bridging Award, Brookhaven National Laboratory Center for Functional Nanomaterials, and Columbia Nano Initiative, and Office of Naval Research Young Investigator Program.